\documentclass[12pt,a4paper]{article}
\usepackage{amsfonts}

\newtheorem{thm}{Theorem}
\newtheorem{prop}{Proposition}
\newtheorem{lem}{Lemma}

\newtheorem{cor}{Corollary}

 \newcommand{\Real}{\mathbb R}

 \newcommand{\bs}{\bigskip}
 
 \newcommand{\bew}{{\em Proof:}}
 \newcommand{\eb}{\hfill q.e.d.}

\newcommand{\B}{{\rm C}}
\renewcommand{\d}{{\rm d}}

\newcommand{\quadmatrix}[4]{\left(
    \begin{array}{cc} #1&#2\\#3&#4
    \end{array}\right)}

\newcommand{\lie}{\hbox{\rm Lie}\,G}
\renewcommand{\sl}{symplectic leaf}
\newcommand{\sls}{symplectic leaves}
\newcommand{\plg}{Poisson Lie group}
\newcommand{\Dp}{\theta}
\newcommand{\Dm}{\theta^-}
\newcommand{\bp}{B}
\newcommand{\bm}{B^-}
\newcommand{\np}{N}
\newcommand{\nm}{N^-}

\newcommand{\x}{\times}

\renewcommand{\a}{\alpha}
\renewcommand{\b}{\beta}
\newcommand{\dt}{\cdot}
\renewcommand{\d}{\delta}

\renewcommand{\i}{\infty}

\title{\bf Factorization dynamics and Coxeter-
Toda lattices\thanks{This work was partially supported 
by NSF grant DMS-9603239,  DAAD grant ``Integrable systems'' and the
DFG, Sonderforschungsbereich 288}}

\author{\bf{Tim Hoffmann}\addtocounter{footnote}{1}\thanks{timh@sfb288.math.tu-berlin.de}, \bf{Johannes
Kellendonk}\thanks{kellen@math.tu-berlin.de}, \bf{Nadja
Kutz}\thanks{nadja@math.tu-berlin.de}\\
Fachbereich  Mathematik, Sekr. MA 8-5, \\ 
Technische Universit\"at Berlin, \\ Stra\ss e des
  17. Juni 136, \\ 10623 Berlin, Germany\vspace{0.3cm} \\ 
 and {\bf Nicolai Reshetikhin}
\thanks{reshetik@math.berkeley.edu}\\
Department of Mathematics, \\ University of California
  at Berkeley,\\  Berkeley, CA 94720, USA}

\date{\today}

\begin{document}

\bibliographystyle{alpha}
\maketitle

\begin{abstract}
It is shown that the factorization relation on simple Lie groups with
standard Poisson Lie structure restricted to Coxeter symplectic leaves
gives an integrable dynamical system. This system can be regarded as a
discretization of the Toda flow. In case of $SL_n$ the integrals of
the factorization dynamics are integrals of the relativistic Toda
system. A substantial part of the paper is devoted to the study of
symplectic leaves in simple complex Lie groups, its Borel subgroups
and their doubles.
\end{abstract}

\tableofcontents
\section*{Introduction}
\addcontentsline{toc}{section}{Introduction}

An integrable Hamiltonian system on a symplectic manifold consists
of a Hamiltonian that generates the dynamics together with a
Lagrangian fibration on the manifold such that the flow lines
generated by the Hamiltonian are parallel to the fibers. Usually,
the fibers are level surfaces of functions called higher
integrals. The fibration by level surfaces is Lagrangian when the
integrals Poisson commute and the flow lines are parallel to the
fibers when the integrals Poisson commute with the Hamiltonian.

The level surfaces of the integrals are equipped with natural
affine coordinates in which the dynamics is linear \cite{Arnold89}
\cite{HoferZehnder94}. Integrable systems on Poisson Lie groups have the following characteristic features:
\begin{itemize}
\item The phase space of such  system is a symplectic manifold which
is a symplectic leaf of a factorizable Poisson Lie group $G$.
\item The level surfaces of integrals are $G$-orbits with respect to the
adjoint action of the group on itself.
\end{itemize}

One should notice that for some symplectic leaves the
$G$-invariant functions do not form complete set of Poisson
commuting integrals (their level sets are not Lagrangian
submanifolds, but only co-isotropic). In such cases still there is
a complete system of integrals, but the complimentary integrals
may have singularities. An example is so-called full Toda system
\cite{Kostant79} \cite{DLNT86}.

Since symplectic leaves of  Poisson Lie group $G$  are connected
components of orbits of the dressing action of the dual Poisson
Lie group $G^*$ on $G$, the invariant tori of such systems lie in
the intersection of $Ad_G$ and $G^*$-orbits in $G$.

Surprisingly enough, most of the known integrable systems on
Poisson Lie groups are of this type. Such integrable systems have
a Lax representation. Systematic treatment of such integrable
systems was done by Semenov-Tian-Shanskii \cite{STS85}.
Linearization of this construction in a neighborhood of identity
gives the similar construction based on Lie algebras which has
been pioneered by Kostant \cite{Kostant79} on the example of Toda
lattices and by Adler \cite{Adler79} on the example of KdV
equation.

An integrable discrete dynamical system on a symplectic manifold is a
symplectomorphism which acts parallel to fibers of a Lagrangian
fibration given by level surfaces of integrals.  More generally,
it can be a Poisson relation preserving the fibration, for details
see \cite{Veselov91}.

In this paper we derive integrable systems related to Toda
models \cite{Toda88} (and references therein). 
We show that for simple Lie groups the
factorization relation restricted to symplectic leaves
that are associated with a Coxeter element in the Weyl group 
yields a discrete integrable evolution. Such a dynamical system will be
called {\em Coxeter-Toda lattice} and the dynamics {\em factorization 
dynamics}. It
turns out that different choices of Coxeter element produce 
isomorphic integrable systems.
The integrals for the factorization dynamics are in case of 
 $G=SL_n$ the integrals of so-called relativistic Toda
lattice introduced in \cite{Ruijsenaars90}.
(Since we will deal only
with simple Lie groups with standard Poisson Lie structure we 
can avoid going into the general discussion of factorizable Poisson
Lie groups.)
The phase space of the Coxeter-Toda lattice the symplectic leaf 
mentioned above.
On a Zariski open subset of such a leaf 
which is isomorphic to ${\Bbb C}^{2r}$  
one can introduce coordinates $\chi^{\pm}_i$, $i=1,\dots,r=\mbox{\rm rank}
G$ with the following Poisson brackets:
\begin{eqnarray*}
\{\chi_i^{\pm}, \chi_j^{\pm}\} &=& 0 \\
\{\chi_i^+, \chi_j^-\}& =& - 2 d_i \B_{ij} \chi_i^+ \chi_j^-
\end{eqnarray*}
Here $\B_{ij}$ is the Cartan matrix of $G$ and the $d_i$ co-prime
positive integers symmetrizing it.

The factorization relation restricted to a Coxeter symplectic leaf
gives a symplectomorphism which acts on coordinates $\chi_i^{\pm}$
is as follows
\begin{eqnarray*}
\alpha(\chi_i^+) &=&\chi_i^-\\ \alpha(\chi_i^-) &=&
\frac{(\chi_i^-)^2}{\chi_i^+} \prod_{j=1}^r (1 -
\chi_j^-)^{-\B_{ji}}.
\end{eqnarray*}
This symplectomorphism is integrable. We will call it the discrete
Toda evolution. Its integrals have the following description in terms
of characters of finite dimensional representations of $G$.

Let $x^-_i, h_i, x^+_i$ be Chevalley generators of the Lie algebra
${\frak g}= Lie (G)$ and $\varphi_i: SL_2(i)\subset G$ be the
natural embedding of the $SL_2$ subgroup generated by the elements
$x^-_i, h_i, x^+_i$ corresponding to the simple root $\alpha_i$.
For a Coxeter element $w$ of the Weyl group $W$ of $G$ fix a
reduced decomposition $w=s_{i_1}\dots s_{i_r}$ where
$r=rank({\frak g})$ and define the element of $G$
$$
g=\prod_{j=1}^r(\frac{\chi^+_j}{\chi^-_j})^{h^j}
exp(-\chi^+_{i_1}x^+_{i_1})exp(x^-_{i_1})\dots
exp(-\chi^+_{i_r}x^+_{i_r})exp(x^-_{i_r}) $$
Here $\{h^j\}_{j=1}^r$ are elements of the Cartan subalgebra  of
$\frak g$ corresponding to fundamental weights,
$h_i=\sum_{j=1}^r \B_{ij}h^j$.

The functions
\begin{equation}\label{char}
Ch_V(\chi^+,\chi^-)=Tr_V(g)
\end{equation}
where $V$ is a finite dimensional representation of $G$ form
Poisson commutative subalgebra in the algebra of functions the
phase space. They are the integrals of the map $\alpha$. The
characters of fundamental irreducible representations of $G$
generate the subalgebra of integrals.

Consider the function $$ H_d(\chi^{\pm})= \frac{1}{2} (\xi,\xi)$$
where $g=exp(\xi)$ and $(.,.)$ is the Killing form on $\lie$. The
Hamiltonian flow generated by this function interpolates the
map $\alpha$.

For $G=SL_n$ the integrals (\ref{char}) are the integrals of
so-called relativistic Toda lattice \cite{Ruijsenaars90}. In a
neighborhood of the identity these integrals turn into the
integrals of the (usual) Toda lattice. In the same sense as a Lie
algebra can be regarded as a linearization of a Lie group, the
usual Toda lattices are linearizations of Coxeter-Toda lattices . 

Integrable discretizations of Toda lattices  have been discovered
by Hirota \cite{Hirota77} who studied solitonic aspects of
them (see also \cite{DateJimboMiwa82}). Later they were re-derived in
\cite{Suris90,Suris91b} from discrete time version of a Lax pair.
The Hamiltonian interpretation based on classical $r$-matrices was
derived in \cite{Suris91a} and generalized to Toda systems related
to all classical Lie groups (and their affine extensions). In
\cite{KashaevReshetikhin97} a discrete version of Toda field theory
was described together with the Hamiltonian structure and its
quantization.

The role of matrix factorization in discrete integrable systems
was noticed quite some time ago. The references include
\cite{Symes82}\cite{QuispelNijhoffCapelVanderLinden84} \cite{MoserVeselov91}
\cite{DeiftLiTomei89}.

The primary goal of this article is not to produce new discrete
integrable systems (although those related to exceptional Lie
groups are new) but rather to demonstrate how the discrete Toda
evolution together with its integrals (\ref{char}) can be
derived in a systematic way from the geometry of Poisson Lie
groups, and from the factorization relation.

A large part of this paper is devoted to the study of the phase
space of these systems. This requires the careful study of
symplectic leaves of $B$ (a Borel subgroup in a simple algebraic
Lie group $G$ with the standard Poisson Lie structure) and of its
double.

 In section 1 we remind basic facts about Poisson Lie
groups and describe the factorization dynamics on factorizable
Poisson Lie groups. Section 2 contains the analysis of symplectic
leaves of simple complex algebraic groups $G$ with a standard
Poisson Lie structure. In section 3 we describe symplectic leaves
of the Borel subgroup $B$ of a simple Poisson Lie group $G$.
Section 4 contains the description of symplectic leaves of the
double of $B$ and of how they are related to symplectic leaves of
$B$ and of $G$. The factorization dynamics on Coxeter symplectic
leaves is studied in section 5. The interpolating flow and the
relation to the (usual) Toda lattices is described in section 6.
In the conclusion we point out what may be done next in
this direction.

\section{Basic facts about simple Poisson Lie groups}
\label{SymplecticLeaves}

\subsection{Basic facts about Poisson Lie groups}

A Poisson Lie group is a Lie group equipped with a Poisson
structure which is compatible with the group multiplication.

There is a functorial correspondence between connected, simply
connected \plg s and Lie bialgebras \cite{Drinfeld87}. The Lie
bialgebra corresponding to a given Poisson Lie group is called
tangent Lie bialgebra.

The dual of a Lie bialgebra $p$ is the dual vector space $p^*$
equipped with the Lie bracket dual to Lie cobracket of $p$ and
with the Lie cobracket dual to Lie bracket on $p$. The dual $P^*$
of a \plg\ $P$ is, by definition, the connected, simply connected
\plg\ having the dual $p^*$ of the Lie bialgebra $p$ corresponding
to $P$ as Lie bialgebra. Denote by ${p^*}^{op}$ the Lie bialgebra
$p^*$ with opposite cobracket (which is minus the original
cobracket).

The double $D(p)$ of $p$ is the direct sum $p\oplus {p^*}^{op}$ as
a Lie coalgebra and its Lie bracket is determined uniquely by
the requirement that the natural inclusions $i:p\to D(p)$ and $j:
{p^*}^{op}\to D(p)$ (into the first and second summand,
respectively) are Lie bialgebra homomorphisms.

The double $D(P)$ of $P$ is the connected, simply connected
Poisson Lie group having $D(p)$ as its Lie bialgebra. The maps $i$
and $j$ lift to injective Poisson maps $i:P \to D(P)$,
$j:{P^*}^{op} \to D(P)$ and consequently to a map
$\mu\circ (i\times j):P\times {P^*}^{op} \to D(P)$: $(x,y)\mapsto
i(x)j(y)$ which is also a local Poisson isomorphism. By a local
isomorphism we mean an isomorphism between neighborhoods of the
identity.

A \sl\ of a Poisson manifold is an equivalence class of points
which can be joined by piecewise Hamiltonian flow lines. When the
Poisson manifold is a \plg\ $P$, there is another description of
these leaves which involves the dressing action of the dual
Poisson Lie group on $P$. 

The Poisson Lie group $P^*$ acts on $D(P)$ via left
multiplication, $y\cdot x:= j(y)x$. We also have a map $\varphi
:P\to D(P)/j(P^{*op})$ which is the composition of $i$ with the
natural projection. In a neighborhood of the identity this map
$\varphi$ is a Poisson isomorphism and induces dressing action of
$P^*$ on $P$ \cite{STS85}.The map $\varphi$ is a finite cover and has open
dense range. 

\newcommand{\dc}{double-coset}
The \sls\ of $P$ are orbits of dressing action of $G^{*op}$ and are
connected components of preimages
of left $P^*$-orbits in $D(P)/j(P^{*op})$. 

Among the cases which have been investigated we point out the
following three, $P=G$ (a complex connected and simply connected
simple Lie group with standard Poisson structure), $P=B$ a
Borel-(Poisson)-subgroup of $G$, and $P=K$ the compact real from
of $G$.

For $P=K$, the double, which can be identified with $G$ as a real
group, is globally isomorphic to $K\times {K^*}^{op}$ as a real
manifold via Iwasawa factorization. The map $\varphi$ in this case
is a global Poisson isomorphism \cite{ LuWeinstein90}. There is
particular simple relation between the Bruhat decomposition of $K$
and its \sls\ \cite{Soibelman90,LuWeinstein90}. It is worth
noticing that as the double of $K$ the complex simple Lie group is
equipped with real Poisson structure which is different from the
standard Lie Poisson structure on $G$.

In the first two cases, which are the ones we shall consider in
detail below, the double is only locally isomorphic to $P\times
{P^*}^{op}$. The symplectic leaves
of $G$ have been studied in
\cite{HodgesLevasseur93} .
Symplectic leaves for $B$ were
described in \cite{DeConciniKacProcesi95}. We reproduce the
results of \cite{HodgesLevasseur93}  \cite{DeConciniKacProcesi95}
below but will describe symplectic leaves in $G$ and $B$
more explicitly. 

\newcommand{\gm}{G^{-}}
\newcommand{\jg}{j(\gm)}

\subsection{Standard Poisson structure on a simple Lie group}
Let $G$ be a simple complex Lie group.
Fix a labeling of the nodes on the Dynkin diagram associated
with the Lie algebra $\lie$ by integers $i=1,\dots, r=rank(G)$.
Assign the simple root $\alpha_i$ to the node labeled by $i$ .
Let $\B$ be the Cartan matrix, that is,
$$\B_{ji}=2\frac{(\alpha_i,\alpha_j)}{(\alpha_j,\alpha_j)}.$$
Denote by  $d_i$ the length of $i$-th simple root, then
$d_i\B_{ij} = d_j\B_{ji}$.

Fix a Borel subgroup $B \subset G$. This fixes the polarization of
the root system and together with the enumeration
of nodes of Dynkin diagram fixes the generators
of the Lie algebra $\lie$ $\{h_i,x_i^{\pm}\}_{i=1,\cdots,r}$
corresponding to simple roots of $\lie$. The determining relations
for these generators are:
\begin{eqnarray*}
[h_i,h_j] &=& 0, \\
{[x_i^+,x_j^-]} & = & \delta_{ij}h_i, \\
{[h_i,x_j^{\pm}]} &=&\pm \B_{ij}x_j^{\pm} \\
ad(x_i^{\pm})^{1-\B_{ij}}x_j^{\pm}&=&0, \quad i\neq j.
\end{eqnarray*}

The standard Lie bialgebra structure on $\lie$ compatible with the chosen
Borel subgroup $B$ is given by the cobracket
acting on generators as follows
$$
\begin{array}{lr}
\delta(h_i) = 0,&\delta(x_i^\pm) = d_i x_i^\pm\wedge h_i.
\end{array}
$$

This induces the Poisson Lie structure on $G$ for which the Lie
bialgebra described above is the tangent Lie bialgebra. The Borel
subgroup $B$ and its opposite $B^-$ are Poisson Lie subgroups.

The Lie bialgebra $Lie(G)$ is isomorphic to the double of the Lie
bialgebra $Lie(B)$ quotioned by the diagonally embedded Cartan
subalgebra \cite{Drinfeld87}.

We denote by $\np$ and $\nm$ the nilpotent subgroups of $\bp$ and
$\bm$, respectively. Since $H=\bp\cap\bm$ we have two natural
projections and isomorphisms $\Dp:\bp\to \bp/\np\cong H$ and
$\Dm:\bm\to \bm/\nm\cong H$. We shall also write $B^+$ and $N^+$ for
$B$ and $N$, respectively.

\newcommand{\qed}{\eb}

\newcommand{\e}{\varepsilon}
\newcommand{\G}{\Gamma}

\section{Symplectic leaves of $G$}

\subsection{Bruhat decomposition of the double of G}
A simple Lie group $G$ with fixed
Borel subgroup $B$ admits Bruhat decomposition with
respect to $B$:
$$
G=\bigcup_{w\in W} BwB
$$
Here $BwB \stackrel{\mbox{def}}{=} B\dot wB$ where
$\dot w$ is a representative of $w\in N_G(H)/H$ in
$N_G(H)$ (clearly $B\dot w B$ depends only on the
class $w\in N_G(H)/H$).

There is also a Bruhat decomposition of $G$ with
respect to $B^-$:
$$
G=\bigcup_w B^-wB^- \ .
$$
Recall \cite{KorogodskiSoibelman98} that the double $D(G)$ is,  as a group, 
isomorphic to $G\x G$.
The cell decompositions of $G$ therefore give the
Bruhat decomposition of $D(G)$  with respect to
$D^-=B^-\times B$:
$$
D(G)=\bigcup_{(w_1,w_2)\in W\x W} D^-(w_1,w_2)D^-,
$$
$D^-(w_1,w_2)D^-=B^-w_1B^-\x Bw_2B$,
where $W\x W=N_{D(G)}(H\x H)/H\x H$ is the Weyl group of
$D(G)$. We can also represent $D^-\subset D(G)$ as
$$
D^-=(H\x H)(N^-\x N^+)= (N^-\x N^+)(H\x H) \ .
$$
Then for the Bruhat cell $D^-(w_1,w_2)D^-$ we can
write
\begin{equation}\label{e1}
D^-(w_1,w_2)D^- = (N^-_{w_1}\x N^+_{w_2})(H\x H)
(\dot w_1,\dot w_2)D^-
\end{equation}
where $N^{\pm}_w=\{n\in N^{\pm} | \dot w^{-1}n\dot
w\in N^{\mp}\}$ (clearly this definition of $N^{\pm}_w$ does not depend on
the choice of $\dot w$).

\subsection{Left cosets $D(G)/j(G^-)$}
Let $G^-=G^{*op}$ which may be identified with
$\{ (b^-,b)\in B^-\times B|\theta^-(b^-)=\theta(b)^{-1}\}$,
a subgroup of  $B^-\times B$,
\cite{KorogodskiSoibelman98}. 
We write $j:G^-\hookrightarrow B^-\times B$ for this identification.
There is a natural isomorphism:
\begin{equation}
D^-/j(G^-)\simeq H . \label{RA1}
\end{equation}
The group $H\x H$ acts on cosets $(\dot w_1,\dot w_2)D^-/j(G^-)$
by left multiplication:
\begin{eqnarray} (h,h')(\dot w_1,\dot w_2)(b^-,
b)j(G^-)  &=& (\dot w_1,\dot w_2) (h_{w_1}b^-,h'_{w_2}b)j(G^-)
\nonumber \\&=&(\dot w_1,\dot
w_2)(Ad_{h_{w_1}}b^-,h'_{w_2}h_{w_1}(Ad_{h_{w_1}^{-1}}b)j(G^-)
\nonumber \\ &=& (\dot w_1,\dot
w_2)(\theta^-(b^-),h'_{w_2}h_{w_1}\theta(b))j(G^-)\  \nonumber .
\end{eqnarray}
Here $(b^-,b)\in B^-\times B$ and we write $h_w=\dot w^{-1}h\dot w$. 
Using also (\ref{RA1}) 
we conclude that this action has
stationary subgroup
\begin{equation}
H^{w_1,w_2} = \{ (h,h')\in H\x H\mid h_{w_1}= {h'_{w_2}}^{-1}\}\ . \label{RA2}
\end{equation}
Thus, we have an isomorphism $D^-(w_1,w_2)D^-/j(G^-)\cong
N^-_{w_1}\times N^+_{w_2}\times H$ and, in particular,
$dim(D^-(w_1,w_2)D^-/j(G^-))=l(w_1)+l(w_2)+r$.

\subsection{Double cosets $j(G^-)\backslash D(G)/ j(G^-)$}

For double cosets, we have $$ j(G^-)(n^-_{w_1},
n^+_{w_2})(h_1,h_2) (\dot w_1,\dot w_2)(b^-,b)j(G^-) = j(G^-)
(\tilde{h}_1,\tilde{h}_2) (\dot w_1,\dot w_2)j(G^-) $$ where
$\tilde{h}_1=h_1\theta^-(b^-)_{w_1^{-1}}, \tilde{h}_2=
h_2\theta^-(b^-)_{w_2^{-1}}$.
The set of such double cosets is according to (\ref{RA2}) isomorphic
to
\begin{equation} \label{co}
j(H)\backslash H\x H/j_{w_1w_2}(H)
\end{equation}
where $j(H)\subset H\x H$ is the subgroup that consists of
elements $(h,h^{-1})$, $h\in H$ and
$j_{w_1,w_2}(h)=(h_{w_1},h_{w_2}^{-1})$.
The coset of $(h_1,h_2)\in H\times H$ in (\ref{co}) is the set 
$\{(h,h^{-1})(1, h_1h_2h''{h}_{w_2^{-1}w_1}^{\prime\prime\,-1})|h,h''\in
H\}$. Thus (\ref{co}) is isomorphic to $H_{w_2^{-1}w_1}$, 
where $H_w$ is the space of
$H$-orbits on $H$ with respect to the action
\begin{equation}\label{e2}
\label{actio} h:h'\to h'hh_{w}^{-1}.
\end{equation}
All orbits are naturally isomorphic and we denote the one through $1$
by $H^w$. Furthermore, $H_w$ is isomorphic to
$\ker{(w_2^{-1}w_1-id)}=\{h\in H|h_{w_2^{-1}w_1}=h\}$. 
Thus, we proved:
\begin{prop} \label{double} We have an isomorphism
\begin{equation}
j(G^-)\backslash D^-(w_1,w_2)D^-/ j(G^-) \simeq H_{w_2^{-1}w_1}.
\label{RA3}
\end{equation}
Each \  $j(G^-)$ orbit corresponding to an element of this set is
isomorphic to 
\begin{equation}
N^-_{w_1}\x N^+_{w_2}\x H^{w_2^{-1}w_1}
\label{RA4}
\end{equation}
In particular, each such orbit has the dimension $$
\ell(w_1)+\ell(w_2)+\mbox{dim}(\mbox{coker}(w_2^{-1}w_1-1)) .$$
\end{prop}
Notice that the isomorphism (\ref{RA3}) 
and the isomorphisms between $j(G^-)$-orbits and sets (\ref{RA4}) are not
canonical but depend on the choice of representatives $\dot
w_1,\dot w_2$. What we really have here is a fiber bundle
\begin{equation} \label{RA5}
D^-(w_1,w_2)D^-/j(G^-) \to j(G^-)\backslash D^-(w_1,w_2)D^-/j(G^-)
\end{equation}
over the torus $H_{w_2^{-1}w_1}$ whose fibers are $j(G^-)$-orbits.

\subsection{Symplectic leaves of $G$ and double Bruhat cells}

\label{slg} Double Bruhat cells are defined as
intersections of $B$-Bruhat
cells and $B^-$-Bruhat cells: $$ G^{w_1,w_2} = B^-w_1B^- \cap
Bw_2B \ . $$ It is known that $dim(G^{w_1,w_2})=l(w_1)+l(w_2)+r$
(for example \cite{FominZelevinsky99}).

 Let $\varphi:
G \stackrel{i}{\hookrightarrow} D \to D/j(G^-)$ be the composition
of diagonal embedding with the natural projection. According to 
(\ref{e1}) we have 
$$ D^-(w_1,w_2)D^-/j(G^-) \cong (N^-_{w_1}\x
N^+_{w_2})(\dot w_1,\dot w_2) i(H) \ . $$
Define $$ \Gamma:=\{ \varepsilon\in H|
\varepsilon^2=1\} $$.
\begin{thm}
The restriction of $\varphi$ to
$G^{w_1,w_2}\to D^-(w_1,w_2)D^-/ j(G^-)$ is a
cover map with group of deck transformations $\Gamma$.
Its image is a dense open subset .
\end{thm}
Here is the outline of the proof. Let $g=n^-_{w_1}{\dot
w_1}b^-=n^+_{w_2}{\dot w_2}b^+\in B^-w_1B^- \cap Bw_2B$
where $n^-_{w_1}\in
N^-_{w_1}, \ n^+_{w_2}\in N^+_{w_2}$ and $b^{\pm}\in B^{\pm}$. Then we
have $$ \varphi (g)=(g,g)j(G^-)=(n^-_{w_1}{\dot
w_1}b^-,n^+_{w_2}{\dot w_2}b^+)j(G^-). $$ 
Therefore $\varphi(g)$ is an element of $D^-(w_1,w_2)D^-/ j(G^-)$.

Conversely, assume $x_1=n^-_{w_1}{\dot w_1}b^-$ and
$x_2=n^+_{w_2}{\dot w_2}b^+$ then $(x_1, x_2)j(G)$. This class has
a representative of the form $(g,g)j(G)$ if and only if there
exists $(\eta_+,\eta_-)\in G_-$ such that $n^-_{w_1}{\dot
w_1}\eta_-=n^+_{w_2}{\dot w_2}\eta_+$. According to 
\cite{FominZelevinsky99} 
such elements exist on a dense open subset of $N^-_{w_1}
\times N^+_{w_2}$. Therefore the image of $\varphi$ is open
dense in $D^-(w_1,w_2)D^-/ j(G^-)$. Furthermore 
$$ \varphi(g\varepsilon)=(g\varepsilon,
g\varepsilon^{-1})j(G^-)=\varphi(g) $$ for each $\varepsilon\in\Gamma$.
This shows that $\Gamma$ acts fixed point freely on the preimages of points.
Since $i(\Gamma) = i(H)\cap j(H)$ 
is the kernel of $\varphi$, $\Gamma$ is the group of deck transformations. 

Since the symplectic leaves in $G$ are connected components of
preimages of $j(G^-)$-orbits in $D(G)/ j(G^-)$ we obtain the
following description of leaves.
\begin{cor} The double Bruhat cell $G^{w_1,w_2}$ is a collection of
symplectic leaves of $G$ each one being a connected component of the
preimage of a double coset $j(G^-)\backslash D^-(w_1,w_2)D^-/ j(G^-)$.
\end{cor}
We will describe symplectic leaves which belong to
$G^{w_1,w_2}$ more explicitly later using Hamiltonian reduction.

\section{Symplectic Leaves of $B$}

\subsection{$B^-$ double cosets in $D(B)$}

The double of a Borel subgroup $B$ of $G$ is
isomorphic to $G\x H$ as a group \cite{KorogodskiSoibelman98}. 
Furthermore, $B^{*op}\cong B^-$, sitting inside $G\x H$ as
$\underline{j}:B^-\to G\x H$: $\underline{j}(b^-)=(b^-,\Theta^-(b^-))$.
In particular, $D(B)$ has the following cell decompositions
\begin{equation} \label{brd}
D(B)=G\x H=\bigcup_{w\in W} B^-w B^-\x H=\bigcup_{w\in W} Bw B\x H
\ .
\end{equation}
Denote $D(B)_w=B^-wB^-\times H$ and $D(B)^w=BwB\times H$.
For the quotient $D(B)/\underline{j}(B^-)$ we have $$
D(B)/\underline{j}(B^-) = \bigcup_{w\in W} (B^-w B^-\x
H)/\underline{j}(B^-)\cong \bigcup_{w\in W} N^-_w \x H .\ $$ 
Let us compute double cosets:
\begin{eqnarray}
\underline{j}(B^-)(b^-{\dot w}\tilde b^-,h) \underline{j}(B^-) &
=& \underline{j}(B^-)(hb^-\dot w\tilde b^-,1) \underline{j}(B^-)
\nonumber \\ &=&\underline{j}(B^-)(hh^-\dot w\tilde h^-,1)
\underline{j}(B^-) \nonumber \\ &=&\underline{j}(B^-)(\dot
wh',1)\underline{j}(B^-)\nonumber \\ &\simeq&
\underline{j}(H)(\dot w h',1)\underline{j}(H) \nonumber \ .
\end{eqnarray}
Clearly $\underline{j}(h)=(h,h^{-1})\in H\times H\subset G\times H$.
Therefore we have the isomorphism $$ \underline{j}(B^-)\backslash
D(B)_w/ \underline{j}(B^-)\cong H_w $$ where, we recall, 
$H_w$ is the space of $H$-orbits on $H$ for the action 
(\ref{e2}). In particular, $dim(H_w)=dim(ker(w-id))$.

Choose a point $(n^-_w\dot w,h)$ in $D(B)$ representing an
equivalence class in $D(B)/\underline{j}(B^-)$. The left
$\underline{j}(B^-)$ orbit passing through this point is the set
of elements
\begin{eqnarray}
&&\{ (b^-n^-_w\dot w,\theta(b^-)^{-1}h)\underline{j}(B^-),
 b^-\in B^-\} \nonumber \\
&=&
\{(\tilde n^-_w\dot w\theta(b^-)_w,\theta(b^-)^{-1})h)
\underline{j}(B^-)\mid b^-\in B^-, \ b^-n_w^- =\tilde n^-_w
\theta(b^-)\} \nonumber \\
&=&\{ (\tilde n^-_w\dot w,\theta(b^-)^{-1}\theta (b^-)_wh)
\underline{j}(B^-)\mid b^-\in B^-, \ b^-n_w^- =\tilde n^-_w
\theta(b^-)\} \nonumber
\end{eqnarray}
Thus, this orbit is isomorphic to 
$$ N^-_w\x H^w.$$ 
Finally, since these orbits are isomorphic to
$({\Bbb C})^{\ell(w)}\times ({\Bbb
C}^{\times})^{dim(coker(w-id))}$  we proved the theorem:

\begin{thm} Each  $\underline{j}(B^-)$-orbit in
$D(B)_w/\underline{j}(B^-)$ is isomorphic to $({\Bbb
C})^{\ell(w)}\times ({\Bbb C}^{\times})^{dim(coker(w-id))}$.
\end{thm}

Similar to the case of $G$, the isomorphisms are
not canonical but we have a fiber bundle
$D(B)_w/\underline{j}(B^-)\to
\underline{j}(B^-)\backslash D(B)/\underline{j}(B^-)$ 
whose fibers are the $\underline{j}(B^-)$ orbits.

\subsection{Factorization of left  cosets}

For $w\in W$ define the subset $B_w=B\cap B^-wB^-$.
Fix a reduced decomposition $$ w=s_{i_1}\dots s_{i_{\ell(w)}} $$
where $\ell(w)$ is the length of $w$. Consider the subset $$
B_{i_1,\dots, i_{\ell(w)}} =B_{s_{i_1}}\dots
B_{s_{i_{\ell(w)}}}\subset B $$ where $B_{s_i}=B(i)\cap
B(i)^-s_iB^-(i)$ and $B(i)=B\cap SL_2(i)$ is the intersection of
the Borel subgroup in $G$ and of the $SL_2$-subgroup generated by
the $i$-th simple root. The set $B_{i_1,\dots, i_{\ell(w)}}$ is
the image of $B_{i_1}\x\dots\x B_{i_{\ell(w)}}$ under
the multiplication in $G$.

For $w\in W$ define numbers of ``repetitions" $$ n_i=\{\# \
{\mbox{ of $i$ in the sequence}} \ \{i_1,\dots ,i_{\ell(w)}\}\} $$
and define the support of $w$ as $I(w)=\{i\mid 1\leq i\leq r, \
n_i\neq 0\}$.

If $n_i\geq 1$ consider the following action of $(\Bbb
C^{\x})^{n_i-1}$ on $B_{i_1}\x\dots\x B_{i_{\ell(w)}}$:
\begin{eqnarray}
&&(x_1,\dots ,x_{n_i-1}): (b_{i_1},\dots ,b_{i_{\ell(w)}})\mapsto
\nonumber \\ && (\dots ,b_i\varphi_i(x_1),\dots
,{\mbox{Ad}}_{\varphi_i(x_1)} (b_j),\dots ,
\varphi_i(x_1)^{-1}b_i{\varphi_i(x_2)},\dots \nonumber \\ &&
{\mbox{Ad}}_{\varphi_i(x_2)} (b_k),\dots ,
\varphi_i(x_2)^{-1}b_i\varphi_i(x_3),\dots )
 \label{RB2}
\end{eqnarray}
Here $\varphi_i: \Bbb C^{\x}\hookrightarrow SL_2\hookrightarrow G$
is the composition of embedding, $\Bbb C^{\x}$ into $SL_2$ as the
(complex) Cartan subgroup and $SL_2$ into $G$ as the $i$-th
$SL_2$-triple. It is clear that for different $n_i,n_j$, both greater
than $1$, the corresponding actions commute so that $w$ gives rise to
an action of the torus $J$, the product of all $(\Bbb
C^{\x})^{n_i-1}$, over $i$ with $n_i>1$. 
\begin{prop} The multiplication map
$B_{s_{i_1}}\x\dots\x B_{s_{i_\ell(w)}} \to B_{i_1,\dots,
i_{\ell(w)}}$ commutes with the $J$-action, assuming $J$ acts
trivially on $B_{i_1,\dots, i_{\ell(w)}}$ and establishes a
birational isomorphism $B_{i_1,\dots, i_{\ell(w)}}\simeq
(B_{s_{i_1}}\times\dots\times B_{s_{i_{\ell(w)}}})/J$.
\end{prop}

Here is the outline of the proof. We can choose the elements for
$(\Bbb C^{\x})^{n_i-1}$ in such a way, that the Cartan parts of
the elements $b_i$ of $(b_{i_1},\dots ,b_{i_{\ell(w)}})$ will all
be trivial, all except one. If will do this for each $i\in I(w)$
we will have cross-section of the action of $J$. Then it quickly
follows  that this cross-section is a birational isomorphism.

The support $I(w)$ of $w$ defines naturally a sub-diagram of the Dynkin
diagram of $G$ (by deleting all nodes not in $I(w)$) and hence a 
subgroup of $G$. Let $B_w'$ be the image in $G$ of the Bruhat cell 
corresponding to $w$ in this subgroup. 
Then multiplication provides an isomorphism
between  $B_w'\times H(w)$ and $B_w$ where
$H(w)$ is the subgroup of 
$H$ corresponding to the simple roots $\alpha_i$ with $i\notin I(w)$. 
The following is known (see for example \cite{FominZelevinsky99}).

\begin{thm} \label{birat}
\begin{itemize}
\item For each $w\in W$ with fixed reduced decomposition the set
$B_{i_1,\dots, i_{\ell(w)}}$ is Zariski open in $B_w'$.
\item For each two reduced decompositions
$w=s_{i_1}\dots s_{i_{\ell(w)}}$ and $w=s_{j_1}\dots
s_{j_{\ell(w)}}$ there is a birational isomorphism between
$B_{i_1,\dots, i_{\ell(w)}}$ and $B_{j_1,\dots, j_{\ell(w)}}$.
\end{itemize}
\end{thm}

\subsection{Symplectic leaves of $B$}
According to the general theory, symplectic leaves of $B$ are
connected components of preimages of $\underline{j}(B^-)$-orbits
in $D(B)/\underline{j}(B^-)$ with respect to the map $\varphi :
B\subset D(B) \to D(B)/\underline{j}(B^-)$.

\begin{prop} The restriction of $\varphi$ to $B_w$ is a covering
map $B_w\to D(B)_w/\underline{j}(B^-)$. The corresponding group of deck
transformations is isomorphic to $\Gamma$.
\end{prop}

Let us describe the symplectic leaves of $B_w$ more explicitly, 
using results of previous subsection.

There is a natural coordinate system in a neighborhood of the
identity of the subgroup $B(i)$ in which the groups elements are
written as $$
\exp{(a_ih_i+b_ix^+_i)}=\exp(a_ih_i)\exp(b'_ix^+_i),$$ where
$b'_i=e^{-a_i}\frac{b_i}{a_i} \sinh(a_i)$.

The corresponding global coordinates on $B(i)$ are $A_i=e^{a_i}, \
B_i=b_i\frac{\sinh(a_i)}{a_i}$. In these coordinates the above element
is represented by the $2\times2$ matrix
$\quadmatrix{A_i}{B_i}{0}{A_i^{-1}}$ in two dimensional
representation of $SL_2$.

The subgroup $B(i)$ is a Poisson Lie subgroup in $SL_2(i)$ with
the following Poisson brackets between coordinate functions $$
\{A_i,B_i\}=-d_iA_iB_i \ . $$ Here and below we will abuse
notations and will denote coordinates and coordinate functions by
the same letters.

The symplectic leaves of $B(i)$ are, one 2-dimensional leaf
$B_{s_i}=\{A_i,B_i\mid A_i\in\Bbb C^{\x}, \ B_i\in\Bbb C^{\x}\}$,
and a 1-dimensional family of zero-dimensional leaves $\{A_i=t,
\ B_i=0\}$.

The product $B_{s_{i_1}}\x\dots\x B_{s_{i_\ell(w)}}$ carries
natural product symplectic structure. Since the multiplication map
is Poisson, the sub-manifold $B_{i_1,\dots, i_{\ell(w)}}\subset
B_w'$ is a Poisson sub-manifold. According to Theorem~\ref{birat},
$B_{i_1,\dots, i_{\ell(w)}}$ is Zariski open in $B_w'$ which implies that
the symplectic leaves of $B_{i_1,\dots, i_{\ell(w)}}$ are
Zariski open sub-varieties in the symplectic leaves of $B$.

The following result, combined with the product Poisson structure on
$B_{i_1} \times \dots \times B_{i_{\ell(w)}}$ allows to describe
symplectic leaves of $B_{i_1,\dots, i_{\ell(w)}}$ explicitly via
Hamiltonian reduction.
\begin{prop} The action (\ref{RB2}) of 
$J=(\Bbb C^{\x})^{\ell(w)-|I(w)|}$ on $B_{i_1}
\times \dots \times B_{i_{\ell(w)}}$ is Hamiltonian.
\end{prop}
Here $|I(w)|$ is the cardinality of the support of $w$. The
Hamiltonians generating this action do not commute. This means
that the pull-back of the moment map maps the Poisson algebra of
functions on $B_{i_1} \times \dots \times B_{i_{\ell(w)}}$ to the
Poisson algebra of functions on a hyper-plane in the vector space
dual to a central extension of Lie algebra $j$. Therefore
symplectic leaves of the quotient space
$(B_{s_{i_1}}\times\dots\times B_{s_{i_{\ell(w)}}})/J$ are
preimages of the corresponding coadjoint orbits with respect to the moment
map. In other words the symplectic leaves of these quotient spaces
can be obtained via Hamiltonian reduction. We will leave the
details of this Hamiltonian reduction to separate publication.

As a corollary of this construction one can derive "product
formulae" for symplectic leaves similar to those derived in \cite{
Soibelman90} for compact Lie groups.

Below we will consider symplectic leaves corresponding to Coxeter
elements. In this case all $n_i=1$ and so $J$ is trivial. It follows
that the product formula doesn't require the Hamiltonian reduction and
a parametrization of the leaf can be given by the coordinates
$A_i,B_i$, $i=1,\cdots,r$.


\subsection{Coxeter symplectic leaves of $B$}
An element of the Weyl groups $W$ is called {\emph Coxeter
element} if its reduced decomposition into the product of  simple
reflections $w=s_{i_1}\dots s_{i_{l(w)}}$ does not have
repetitions in the sequence of sub-indices and if $l(w)=r$ (i.e. in
this product each generator of $W$ appear exactly once).  It is
not difficult to see that if $w$ is a Coxeter element,
$dim(coker(w-id))$ is $r$ and therefore the subset $B_w$ is a
symplectic leaf of $B$. We will call them Coxeter symplectic
leaves.

Let $U_i: B_{s_i} \hookrightarrow B$ be the natural inclusion of
$B_{s_i}\subset B(i)$ into $B$. Then any element of
$B_{i_1,\dots,i_{\ell{w}}}$ can be written as $$
U_{i_1}(A_{i_1},B_{i_1})\cdots U_{i_r}(A_{i_r},B_{i_r}). $$
Thus, for Coxeter symplectic leaves $A_i, B_i$ (more precisely,
their logarithms) are Darboux coordinates.

\subsection{Symplectic leaves of $B^-$}
Symplectic leaves of $B^-$ can be described similarly to how it
was done for $B$. They also can obtained from the ones for $B$
since $B$ is anti-isomorphic to $B^-$ as a Poisson manifold ( there
is an isomorphism of groups, which maps one Poisson tensor to the
negative to the other) .

Let  $C_i, D_i$,  be coordinates on the lower triangular part of
$SL_2(i)$ in which group elements are represented by matrices
$L_i(D_i,C_i)=\quadmatrix{D_i}{0}{C_i}{D_i^{-1}}$ in the two dimensional
irreducible representation of $SL_2$. These coordinate functions
have the following Poisson brackets: $$ \{D_i,C_i\}=d_iD_iC_i. $$
Denote by $B_{s_i}^-$ the sub-variety of the lower triangular part of
$SL_2(i)$ where $C_i\neq 0$.
Fix the Coxeter element $w\in W$ and its reduced decomposition
$w=s_{i_1}\dots s_{i_r}$ . On a Zariski open subset of the Coxeter
symplectic leaf $B^-_w$ one can introduce the natural coordinates
$C_i, D_i$, $i=1,\dots,r$. Every element of this subset can be
written as : $$ L_{i_r}(D_{i_r},C_{i_r}) \dots
L_{i_1}(D_{i_1},C_{i_1}) $$ where $L_i: B^-_{s_i}\subset B^-$ are
natural inclusions.

\section{Symplectic leaves of $D(B)$}

\subsection{Symplectic leaves of $D(B)$}

As above, let us identify $D(B)$ with $G\x H$ as a group. The Poisson
structure on $D(B)=G\x H$ is not the product structure. Symplectic leaves
of $D(B)$ can be described similarly to how it was done for $G$.
Since $D(B)$ is a factorizable Poisson Lie group
\[
D(D(B))\simeq D(B)\x D(B).
\]
Fix this isomorphism together with the identification $D(B)=G\x
H$. This gives the following cell decomposition for $D(D(B))$:
\[
D(B)\x D(B) =\bigsqcup_{w_1,w_2}
(B^-w_1 B^-\x H) \x (Bw_2B\x H) \ .
\]
The Poisson Lie group $D^-(B)=D(B)^{*op}$ can naturally be
identified with $B^-\times B$.

Let $D(B)_w$ and $D(B)^w$ be Bruhat cells of $D(B)$ defined in
(\ref{brd}). The double cosets $j(D^-(B))\backslash
D(B)_{w_1}\times D(B)^{w_2}/j(D^-(B))$ can be computed similarly
to Proposition \ref{double}:
\[
j(D^-(B)) \backslash D(B)_{w_1}\times D(B)^{w_2}/j(D^-(B)) \simeq
H_{w_1}\x H_{w_2}.
\]

The $D^-(B)$-orbit passing through the coset class of $((\dot
w_1,h_1),(\dot w_2,h_2))\in D(B)_{w_1}\x  D(B)^{w_2}$ is
isomorphic to
\begin{equation}\label{(A)}
(N^-_{w_1}\x H^{w_1})\x (N^+_{w_2}\x H^{w_2}).
\end{equation}
Notice that $j(D^-(B))$-orbits in $D(B)_{w_1}\x
D(B)^{w_2}/j(D^-(B))$ are isomorphic to the product of
corresponding orbits for $B$ and for $B^-$.
Again we have a natural fiber bundle
\[
(D(B)_{w_1}\x D(B^{w_2}))/j(D^-(B))\to j(D^-(B))\backslash
D(B)_{w_1}\x D(B)^{w_2}/j(D^-(B))
\]
and $j(D^-(B))$-orbits are fibers of this bundle.
The connected components of preimages of these fibers under
the map $\varphi:D(B)\to 
(D(B)\times D(B))/j(D^-(B))$ are the symplectic leaves of $D(B)$. 
Symplectic leaves whose image
are orbits in $D(B)_{w_1}\x D(B)^{w_2}/j(D^-(B))$ will be denoted
as $S_{w_1,w_2}$.

\subsection{Relation between symplectic leaves of $B$ and $D(B)$ }

Embeddings ${\bold i}:B\hookrightarrow D(B)$ and
${\bold j}:B^-\hookrightarrow D(B)$ combined with the multiplication 
and inversion in
$D(B)$ give rise to the 
map
\begin{equation}
I:B\times B^-\to D(B)=G\times H \ , \qquad
I(b,b^-)=(b(b^-)^{-1},\theta(b)\theta^-(b^-)) \ 
\end{equation}
which is most important to define the factorization
relation, see below.
The image of this map is Zariski open in $D(B)$. This map is also
a Poisson map and therefore maps symplectic leaves of $B\x B^-$ to
symplectic leaves of $D(B)$. The intersection of the image of $I$
 any of the symplectic leaves is Zariski
open in this leaf. This ``explains" the formula (\ref{(A)}).

\subsection{Relation between symplectic leaves of $D(B)$ and $G$}

The Cartan subgroup $H$ acts naturally on $B\x B^-$ by diagonal
multiplication from the right,
\begin{equation}\label{(B)}
h(b,b^-)=(bh,b^-h).
\end{equation}
The following is clear.

\begin{lem}
The map $I$ commutes with the $H$-action
\[
I(bh,b^-h)=I(b,b^-)(1,h^2)
\]
and induces a Poisson map $\tilde I$ between
corresponding cosets:
\[
(B\x B^-)/H \stackrel{\tilde I}{\longrightarrow} G \cong D(B)/H
\]
Here the coset is taken with respect to the action of $H$ on
$D(B)$ by  the multiplication by $(1,h^2)$ from the right.
\end{lem}
It is also clear that the image of $\tilde I$ is Zariski open in
$G$ and that $\tilde I$ is a birational isomorphism. Since the
action (\ref{(B)}) is Hamiltonian, symplectic leaves of $(B\x
B^-)/H$ can be obtained via Hamiltonian reduction from symplectic
leaves of $B\x B^-$. Therefore, symplectic leaves of $G$ can be
obtained via Hamiltonian reduction from symplectic leaves of
$D(B)$.

Symplectic leaves of $G$ can be also described via Hamiltonian
reduction similarly to how it was done for symplectic leaves of
$B$. For this consider two elements $u, v\in W$ and fix their
reduced decomposition $u=s_{i_1}\dots s_{i_l}, v=s_{j_1}\dots
s_{j_m}$. Consider the image 
of $$ B_{s_{i_1}}\times\dots\times B_{s_{i_l}}\times
B^-_{s_{j_m}}\times\dots\times B^-_{s_{j_1}}$$ under the
multiplication and inverse map: $$G_{i_1,\dots,i_l,j_1,\dots,j_m}=
B_{s_{i_1}} \dots B_{s_{i_l}} {B^-_{s_{j_1}}}^{-1}\dots
{B^-_{s_{j_m}}}^{-1}.$$
The double Bruhat cell $G^{u,v}$ has natural decomposition
$G^{u,v}={G'}^{u,v}\times H(u,v)$ where $H(u,v)$ is the subgroup
of $H$ generated by elements corresponding to simple roots which
do not belong to $I(u)\cup I(v)$. It follows from \cite{FominZelevinsky99}
that the  variety $G_{i_1,\dots,i_l,j_1,\dots,j_m}$ is
birationally isomorphic to  ${G'}^{u,v}$. On the other hand it is also
birationally isomorphic to the quotient of $$
B_{s_{i_1}}\times\dots\times B_{s_{i_l}}\times
B^-_{s_{j_m}}\times\dots\times B^-_{s_{j_1}}$$ with respect to the
appropriate Hamiltonian toric action. This allows to construct all
symplectic leaves of $G$ via Hamiltonian reduction. We will leave
the details of this construction for another publication.

\section{Factorization dynamics on Poisson Lie groups}
\subsection{Dynamics of Poisson relations}

Here we will remind basic facts about Poison relations
and their dynamics.
Let $(M,p)$ be a Poisson manifold with the Poisson tensor $p\in
\wedge^2 TM$. Denote by $p^{(2)}\in\wedge^2 T(M\x M)$ the Poisson
tensor corresponding to the following product of Poisson
manifolds: $(M,-p)\x (M,p)$.

A {\it smooth relation} of finite type on a manifold $M$ is a
submanifold $R\subset M\x M$, such that natural projections
$\pi_1,\pi_2:M\x M\to M$, $\pi_1(x,y)=x$, $\pi_2(x,y)=y$ have a
finite number of preimages.

Denote by $T^\perp R$ the forms on $M\x M$ which vanish on
$TR\subset T(M\x M)$.

A smooth relation on $M$ is called a {\it Poisson relation} \ if
\[
p^{(2)}\mid_{T^{\perp}R} =0
\]
and $dim(R)=dim(M)$.

If a relation $R=\{(x,\phi(x))\mid x\in M\}$ is a graph of a map
$\phi: M\to M$ it is Poisson if and only if $\phi$ is a Poisson
map.

An $n$-th iteration of a relation $R$ on $M$ is a submanifold
$R^{(n)}\subset M^{\times (n+1)}$ such that
\[
R^{(n)}=\{(x_1,\dots ,x_{n+1})\mid x_i\in M, \ (x_i,x_{i+1})\in
R\subset M\x M\} \ .
\]
A function $F\in C^\i(M)$ is called an integral of a smooth
relation $R\subset M\x M$ if
\[
F(x)=F(y) \quad {\mbox{for all}} \quad (x,y)\in R \ .
\]
A smooth relation on a symplectic manifold is Poisson if and only
if it is a Lagrangian submanifold in $M\x M$ (equipped with the
product symplectic structure). It is called {\it integrable} if
there exists $n$ independent Poisson commuting functions
$I_1,\dots ,I_n$ which are integrals of $R$.

Similarly one can define Poisson and symplectic relations in an
algebro-geometric setting. For more details about the dynamics of
symplectic relations see \cite{Veselov91}.
\subsection{Factorization relations on Poisson Lie groups}

We will study very specific Poisson relations on Poisson Lie
groups which we will call factorization relations.

Let $P$ be a Poisson Lie group and $D(P)$ be its double. A
factorization relation on $P\x P^{op}$ is a sub-variety ${\mathcal
F}\subset (P\x P^{op})\x (P\x P^{op})$, defined as
\[
{\mathcal F}=\{ (g^+,g^-),(h^+,h^-)\mid
i(g^+)j(g^-)^{-1}=j(h^-)^{-1}i(h^+)\}
\]
where $i:P\hookrightarrow D(P)$ and $j: P^{op}\hookrightarrow
D(P)$  are the natural inclusions of Poisson Lie groups.

\begin{prop} \begin{itemize}\item Functions on $D(P)$ which are invariant with respect to
the adjoint action of $D(P)$ form a Poisson commutative subalgebra
in the Poisson algebra of functions on $D(P)$.

\item A function on $P\x P^{op}$ which is the composition of the map $M(i\x j): P\x
P^{op}\to D(P)$, $(g^+,g^-)\mapsto i(g^+)j(g^-)^{-1}$ and of an
$Ad$-invariant function on $D(P)$ is an integral of the
factorization map.
\end{itemize}
\end{prop}

Part 1 of this proposition is well known \cite{STS85}; part 2 is
obvious:
\[
f(i(g^+)j(g^-)^{-1}) = f(j(h^-)^{-1}i(h^+)) =f(i(h^+)j(h^-)^{-1})
\]

Let $\Sigma_1$ and $\bar{\Sigma_2}$ be symplectic leaves in $P$
and $P^{op}$ respectively. Restricting the relation $\mathcal F$
to the symplectic leaf $\Sigma =\Sigma_1\x\bar{\Sigma_2} \subset
P\x P^{op}$  we obtain a Poisson relation on $\Sigma$ and central
$Ad$-invariant functions on $D(P)$ will produce the integrals of
this relation.

It may happen that one can make a Hamiltonian reduction of
$\Sigma$ in such a way that on the reduced space we have enough
central functions, in a sense that their level surfaces are half
of the dimension of the reduced symplectic manifold. In this case
the factorization dynamics on $\Sigma$ or on the reduced space
will be integrable.

In the next sections we will show that this is exactly what
happens with symplectic leaves corresponding to the Coxeter
elements. As we will see this gives an integrable system which is
a ``nonlinear" version of an open Toda system corresponding to the
Lie algebra $\mathfrak g={\mbox{Lie}}(G)$. It becomes the usual
Toda system in a neighborhood of the identity.

{\bf Remark} One can argue that factorization dynamics is
integrable on all (appropriately reduced) symplectic leaves of
$P\x P^{op}$ when $P$ is a Borel subgroup of simple Lie group $G$.
In a neighborhood of the identity such systems become ``complete"
Toda systems (corresponding to parabolic subgroups in $G$)
\cite{Kostant79} \cite{DeiftLiTomei89}. But this
will be the subject for a separate publication.

\section{Factorization dynamics on Coxeter symplectic leaves}

\subsection{Integrals}

Consider a Coxeter symplectic leaf $S_{w,w}$ of $D(B)$
corresponding to the Coxeter element $w$.  Fix reduced
decomposition $w=s_{i_1}\dots s_{i_r}$. On Zariski open subset
$S'_{w,w}$ of
$S_{w,w}$ each element of $G\cap S_{w,w}$ (provided that $G$ is
embedded to $D(B)=G\x H$ as $(G,e)$) can be represented by the
product
\begin{equation}\label{ulinv} UL^{-1} = U_{i_1}\dots U_{i_r} \
L_{i_1}^{-1}\dots L_{i_r}^{-1} \ .
\end{equation}
Here we abbreviated $U_i\equiv U_i(A_i,B_i)$, $L_i=L_i(D_i,C_i)$.
This
subset depends on the choice of reduced decomposition of $w$. We
will suppress this dependence since different reduced
decompositions give birationally isomorphic subsets.

For each $i=1,\dots ,r$ and given reduced decomposition
$w=s_{i_1}\dots s_{i_r}$ define $\{i\}_+=\{i_\a =1,\dots r\mid
\a >\b, \ i=i_{\b}\}$ \ and \ $\{i\}_-=\{i_\a =1,\dots r\mid \a
<\b, \ i=i_\b\}$.

\begin{prop}
\label{one}   The following identities hold
\begin{eqnarray}\label{A,B}
UL^{-1} &=& \prod_i A_i^{h_i} U_{i_1} (1,{\tilde V}_{i_1})
L_{i_1}^{-1}(1,{\tilde W}_{i_1})\dots U_{i_r}(1,{\tilde
V}_{i_r})\dt
 L^{-1}_{i_r}(1,{\tilde W}_{i_r}) \prod_i D_i^{-h_i} \ ,\nonumber \\
UL^{-1} &=& U_{i_1}(1,V_{i_1})\dots U_{i_r}(1,V_{i_r})\prod_i
\big( \frac{A_i}{D_i}\big)^{h_i} L_{i_1}^{-1}(1, W_{i_1})\dots
L_{i_r}^{-1}(1, W_{i_r})
\end{eqnarray}
where
\begin{eqnarray*}
V_i &=& B_iA_i \prod_{j\in\{i\}_-} A_j^{\B_{ji}} \ , \qquad {\tilde
V}_i = B_iA_i^{-1} \prod_{j\in\{i\}_+} A_j^{-\B_{ji}} \ , \\ W_i
&=& C_iD_i^{-1} \prod_{j\in\{i\}_+} D_j^{-\B_{ji}} \ , \qquad
{\tilde W}_i = C_iD_i \prod_{j\in\{i\}_-} D_j^{\B_{ji}} \ .
\end{eqnarray*}
\end{prop}
The proof of this proposition and of the next lemma is a simple
exercise.

\begin{lem} \ \ ${\tilde V}_i=V_i\prod_j A_j^{-\B_{ji}}$, \
${\tilde W}_i=W_i\prod_j D_j^{\B_{ji}}$.
\end{lem}
Define variables $\chi_i^{\pm}$, $G_i$,  $F_i$ as
\begin{eqnarray*}
\chi^+_i&=&V_i W_i \ , \qquad \chi^-_i=\chi^+_i\prod_j \big(
\frac{A_j}{D_j}\big)^{-\B_{ji}} \\ G_i &=& \frac{B_i}{C_i} A_i D_i 
\prod_{j\in\{i\}_+} A_j^{\B_{ji}}\prod_{j\in\{i\}_-} D_j^{\B_{ji}} ,
\qquad F_i=A_iD_i \ .
\end{eqnarray*}

\begin{prop}\label{two}
 Considered as functions on $S'_{w,w}$,  \
$\chi_i^{\pm}$, $F_i$, and  $G_i$ have the following Poisson
brackets:
\begin{eqnarray*}
\{ \chi^+_i,\chi^+_j\} &=& \{\chi^-_i,\chi^-_j\} =0 \ , \\ \{
\chi^+_i,\chi^-_j\} &=& -2 d_i \B_{ij} \chi^+_i \chi^-_j \ , \\ \{
\chi^\pm_i,F_j\} &=& \{\chi^\pm_i,G_j\}=0 \ , \\ \{F_i,G_j\} &=&
-2d_i F_iG_j \delta_{i,j}\ .
\end{eqnarray*}
\end{prop}

The proof is a straightforward computation based on the definition
of $\chi_i^{\pm}$, $F_i$, and  $G_i$ and on the Poisson brackets
between $A_i$, $B_i$, $C_i$, $D_i$:
\begin{eqnarray*}
\{ A_i,B_j\} &=& -d_i\d_{ij} A_iB_j \ , \\ \{ D_i,C_j\} &=&
d_i\d_{ij} D_iC_j  \ , \\ \{ A_i,A_j\} &=&
\{A_i,C_j\}=\{A_i,D_j\}=\{B_i,C_j\}= \{B_i,D_j\}=\{D_i,D_j\}=0 \
,\\ \{B_i,B_j\}&=&\{C_i,C_j\}=0 \ .
\end{eqnarray*}

Using Proposition \ref{one}, the definition of $\chi^\pm_i$ and
elementary algebra we arrive at the following

\begin{prop}\label{three}
Let $V$ be a finite-dimensional representation of $G$ and
${\rm{Ch}}_V$ be its character. Then
\begin{eqnarray*}
{\rm{Ch}}_V(UL^{-1}) &=& {\rm{Ch}}_V\big( \prod^r_{j=1}
\big(\frac{\chi^+_j}{\chi^-_j}\big)^{h^j} \phi_{i_1}(g_{i_1})\dots
\phi_{i_r}(g_{i_r})\big) \\ &=& {\rm{Ch}}_V\big( \prod^r_{j=1}
\big(\frac{\chi^+_j}{\chi^-_j}\big)^{h^j} \phi_{i_1}({\bar
g}_{i_1})\dots \phi_{i_r}({\bar g}_{i_r})\big)
\end{eqnarray*}
\end{prop}
Here $\phi_i: SL_r(i)\hookrightarrow G$ is the embedding of $SL_2$
generated by $x^+_i$, $h_i$, $x^-_i$ into $G$, \ $g_i$ and ${\bar
g}_i$ are elements of  $SL_2$ whose image in 2-dimensional
irreducible representation is given by the following weight basis
of 2-dimensional irreducible representation:
\[
g_i=\left(\begin{array}{cc} 1\!-\!\chi^-_i & \chi^-_i\\-1 &
1\end{array}\right) \ , \qquad {\bar g}_i=\left(\begin{array}{cc}
1 & \chi^+_i \\-1 & 1\!-\!\chi^+_i\end{array}\right) \ .
\]
The element $\{h^j\}$ forms the basis in ${\frak h}\subset{\frak
g}$ corresponding to fundamental weights: $h_j=\sum_i \B_{ji}h^i$.
Observe that $[h^j,X_i^{\pm}] = \pm \delta_{ij} X_i^{\pm}$, hence
by conjugating $UL^{-1}$ with an element $\exp{a h^i}$ of ${H}$
one can alter the off-diagonals of the $g_i's$. This was used in the 
proof of Proposition \ref{three}.

Now let us interpret these two propositions form the point of view
of Hamiltonian reduction.

\begin{prop}\label{four}
{\rm{(1)}} Functions $\log G_i$ generate $H$ action (\ref{(B)}) on
$S'_{w,w}$.

\noindent {\rm{(2)}} Functions $\log F_i$ generate the adjoint
action of $H$ on $S'_{w,w}\subset D(B)$, \\ $h:(g,h')\mapsto
(hgh^{-1},h')$.
\end{prop}
This proposition can be derived immediately from formulae
(\ref{A,B}) and from the explicit form of Poisson brackets in
terms of coordinates $A_i,B_i,C_i,D_i$.

Characters as functions on the group are invariant with respect to
the adjoint action. Therefore Proposition \ref{four} implies that
characters, computed on $G\cap S'_{w,w}$ do not depend on
$F_i,G_i$ which can be seen also by direct computation
(proposition \ref{three}).

As it follows from 4.2 we can naturally identify
\[
G\cap S'_{w,w}=  S'_{w,w}/H
\] where the $H$ action is
generated by $\log F_i$. Level surfaces of functions $G_i$ are
symplectic leaves of $G\cap S'_{w,w}$ and $\log\chi^\pm_i$ are Darboux
coordinates on these symplectic leaves. All this is clear from the
structure of Poisson brackets in Proposition \ref{two}.

\subsection{Factorization map}

Consider the map \ $\a: ({\Bbb C}^\times)^{2r}\to ({\Bbb
C}^\times)^{2r}$,
\begin{eqnarray*}
\a(\chi^+_i) &=& \chi^-_i \\ \a(\chi^-_i) &=&
\frac{(\chi^{-}_i)^2}{\chi^+_i} \prod_j (1-\chi^-_j)^{-\B_{ji}}\\
\alpha(F_i) &=& F_i\\
\alpha(G_i) &=& G_i \prod_{j} F_j^{\B_{ij}}
\end{eqnarray*}
defined outside of the hyper-planes  ($\chi^-_j=1, \chi^+_i=0$).
Since we are interested in integrable systems whose
Hamiltonians are given by functions on $G$ invariant under 
conjugations and since there functions restricted to a Coxeter
orbit do not depend on $F$ and $G$ variables we will 
focus on the action of the factorization dynamics on $\chi^{\pm}$.
Here we will continue the practice of abusing notation and will denote
coordinates and coordinate functions by the same letter.

Let \ Ch${}_V(\chi^+,\chi^-)$ be functions on $({\Bbb
C}^\times)^{2r}$ as defined in Proposition \ref{three}.

\begin{thm}
${\rm{Ch}}_V(\a(\chi^+),\a(\chi^-)) = {\rm{Ch}}_V(\chi^+,\chi^-)$.
\end{thm}

{\bew} We will use two formulae for these functions derived
in Proposition \ref{three}.
\begin{eqnarray*}
{\rm{Ch}}_V(\a(\chi^+),\a(\chi^-)) &=& {\rm{Ch}}_V\left(
\prod^r_{j=1} \big(\frac{\a(\chi^+_j)}{\a(\chi^-_j)}\big)^{h^j}
\prod^\rightarrow_i \phi_i \left(
\begin{array}{cc} 1 & \a(\chi^+_i)\\-1 & 1\!-\!\alpha(\chi^+_i)\end{array}\right)\right)\\
 &=& {\rm{Ch}}_V\left(
\prod^r_{j=1} \big(\frac{\chi^+_j}{\chi^-_j}\big)^{h^j}
\prod^r_{j=1} (1-\chi^-_j)^{h_j} \prod^\rightarrow_i \phi_i \left(
\begin{array}{cc} 1 & \chi^-_i\\-1 & 1\!-\!\chi^-_i\end{array}\right)\right)\\
 &=& {\rm{Ch}}_V\left(
\prod^r_{j=1} \big(\frac{\chi^+_j}{\chi^-_j}\big)^{h^j}
 \prod^\rightarrow_i \phi_i \left(
\begin{array}{cc} 1 \!-\!\chi^-_i & \chi^-_i\\-1 & 1\end{array}\right)\right)
\end{eqnarray*}
Here the product is taken in the order $(i_1,\dots ,i_r)$. \qed

\begin{prop} Let ${\cal F}\subset S_{w,w}'\x S_{w,w}'$ be the
factorization relation restricted to Coxeter symplectic leaves of
$D(B)$. The diagram
\[
\begin{array}{ccc}
{} & \cal F \\ \chi_L\swarrow & {} & \searrow
\chi_R\vspace{1\jot}\\ ({\Bbb C}^\times)^{2r} &
\stackrel{\a}{\longrightarrow} & ({\Bbb C}^\times)^{2r}
\end{array}
\]
is commutative. Here $\chi_L$ is the composition of the projection
to the first component in $S'_{w,w}\x S'_{w,w}$ and the map $\chi:
(A_i,B_i,C_i,D_i)\mapsto (\chi^\pm_i)$ and $\chi_R$ is the
 composition of the projection to the right
component and $\chi$.
\end{prop}

{\bew} On the image of the factorization map $I: B\times B\to D(B)$,
elements of $S_{w,w^{-1}}\subset D(B)$ can be represented as
\[
(UL^{-1},{\mbox{diag}}(U){\mbox{diag}}(L))
\]
where $U$ and $L$ are as above. The factorization relation $F\subset
S'_{w,w}\x S'_{w,w}$ consists of points
\[
(UL^{-1},{\mbox{diag}}(U){\mbox{diag}}(L)),
({\bar U}{\bar L}^{-1},{\mbox{diag}}({\bar U}){\mbox{diag}}({\bar L}))
\]
satisfying conditions
\begin{eqnarray*}
UL^{-1}&=&{\bar L}^{-1}{\bar U}\\
{\mbox{diag}}(U){\mbox{diag}}(L))&=&
{\mbox{diag}}({\bar U}){\mbox{diag}}({\bar L})
\end{eqnarray*}

Let $U_i,U'_i,U''_i,{\bar U}_i,L_i,L'_i,L''_i,{\bar L}_i$ be factors of
$U,L,\dots$
satisfying relations
\begin{eqnarray*}
UL^{-1}&=& U_{i_1}\dots U_{i_r} L^{-1}_{i_1}\dots L^{-1}_{i_r} =
U'_{i_1}L^{'-1}_{i_1}\dots U'_{i_r}L^{'-1}_{i_r} \\
&=& L^{''-1}_{-1}U''_{-1}\dots L^{''-1}_{i_r}U''_{-r}
= {\bar L}^{-1}_{i_1}\dots {\bar L}^{-1}_{i_r}
{\bar U}_{i_1}\dots {\bar U}_{i_r}
\end{eqnarray*}
Then the coordinates $A,B,C,D$ of these elements have to satisfy the relations
\[
\begin{array}{ll}
A'= A_i \ , & D'_i=D_i\vspace{2\jot}\\ B'_i=B_i\prod_{j\in\{i\}_-}
D_j^{\B_{ji}} & C'_i=C_i\prod_{j\in\{i\}_+} A_j^{-\B_{ji}}
\vspace{2\jot} \\ A'_iD^{'-1}_i-B'_iC'_i =A''_iD^{''-1}_i &
B'_iD'_i=B''_iD_i^{''-1}\vspace{2\jot}\\ C'_iA^{'-1}_i=C''_i A''_i
& A^{-1}_iD_i=D''_i A_i^{''-1}-B''_iC''_i\vspace{2\jot}\\
A'_iD'_i=A''_iD''_i\vspace{2\jot} \\ {\bar
B}_i=B''_i{\displaystyle{\prod_{j\in\{i\}_+}}}D_j^{''\B_{ji}} &
{\bar C}_i=C''_i
{\displaystyle{\prod_{j\in\{i\}_-}}}A_j^{''-\B_{ji}}\vspace{2\jot}\\
{\bar A}_i=A''_i & {\bar D}_i=D''_i
\end{array}
\]
Let us find ${\bar \chi}^+_i$ from these relations:
\begin{eqnarray*}
  {\bar \chi}^+_i 
  &=& {\bar B}_i{\bar C}_i \ \frac{{\bar A}_i}{{\bar
      D}_i} \prod_{j\in\{i\}_-} {\bar A}_j^{-\B_{ji}} \prod_{j\in\{i\}_+}
  {\bar D}_j^{-\B_{ji}}\\ 
  &=& B''_iC''_i \prod_{j\in\{i\}_+}
  D_j^{''\B_{ji}} \prod_{j\in\{i\}_-} A_j^{''-\B_{ji}} \
  \frac{A''_i}{D''_i} \prod_{j\in\{i\}_-} A_j^{''\B_{ji}}
  \prod_{j\in\{i\}_+} D_j^{''-\B_{ji}}\\ 
  &=& B''_iC''_i \
  \frac{A''_i}{D''_i} \
  B'_iC'_i \ \frac{D'_i}{A'_i}\\
  &=& B_iC_i \frac{D_i}{A_i}\prod_{j\in\{i\}_-} D_j^{\B_{ji}}
  \prod_{j\in\{i\}_+} A_j^{-\B_{ji}} \\ 
  &=&\chi_i^+ \frac{\chi^-_i}{\chi^+_i}=\chi^-_i
\end{eqnarray*}
Similarly,
\begin{eqnarray*}
{\bar \chi}^-_i &=& {\bar \chi}^+_i \prod_j \left(\frac{{\bar
D}_j}{{\bar A}_j}\right)^{\B_{ji}} = \ \chi^-_i\prod_j
\left(\frac{D''_j}{A''_j}\right)^{\B_{ji}} \\ &=& \chi^-_i\prod_j
\left(\frac{D'_j}{A'_j} \big( 1-B'_jC'_j
\frac{D'_i}{A'_i}\big)\right)^{-\B_{ji}}\ =\
\frac{(\chi^-_i)^2}{\chi^+_i}\prod_j (1-\chi^-_j)^{-\B_{ji}}
\end{eqnarray*}
Here we used the identities $\prod_j
(D_j'/{A_j'})^{\B_{ji}}=\chi^-_i/{\chi^+_i}$ and
$B_j'C_j'D_j'/{A_j'}=\chi^-_j$. This proves the Proposition.

\begin{cor} The map $\a$ is Poisson.
\end{cor}

This can also be checked by direct calculation using Poisson
brackets between $\chi^\pm_i$.

Thus, we have a Poisson map $\a:(({\Bbb C}^\times)^{2r})\to
(({\Bbb C}^\times)^{2r})$ defined outside of hyper-planes
$\chi^-_i=1, \chi^+=0$, which preserves functions
Ch${}_V(\chi^+,\chi^-)$.

\begin{prop}
{\rm{(1)}} $\{{\rm{Ch}}_V,{\rm{Ch}}_W\}=0$ for every pair of
finite dimensional representations $V$ and $W$.

{\rm{(2)}} ${\rm{Ch}}_V$, as a function of the $\chi^\pm_i$, 
is independent of the choice of the Coxeter element $w$.
\end{prop}

\bew\ 
The first part of this proposition is a general fact about
factorizable Poisson Lie groups.

For the second we have to show that ${\rm{Ch}}_V(\chi^+,\chi^-)$ 
does not depend
on the order $(i_1,\ldots,i_r)$ of the indices. 
Clearly ${\rm{Ch}}_V(\chi^+,\chi^-)$ doesn't
change if we change the order by an elementary transposition (exchange
 of two consecutive indices) of two indices which are
not linked in the Coxeter diagramm. Let us call these transpositions 
free elementary transpositions. Furthermore,
${\rm{Ch}}_V(\chi^+,\chi^-)$
is also invariant under a cyclic permutation as may be seen 
using the observation made after Proposition (\ref{three}). 
Thus the proposition follows from the
easily established fact that every elementary transposition can be 
obtained by successive applications of cyclic permutations and free
elementary transpositions. \eb\bs

To summarize, with each Coxeter symplectic leaf of $G$ we
associated a (complex holomorphic, algebraic) integrable system on
$(({\Bbb C}^\times)^{2r})$ for which the integrals are given by characters
(there are exactly $r$ independent of them) but all these systems 
are trivially isomorphic. The coordinates 
$\chi^\pm_i$ simply describe different points in the group if one changes the
Coxeter element. The factorization
relation restricted to a Coxeter symplectic leaf gives a
discrete-time evolution preserving these integrals.

\subsection{Real positive form}

Consider the real form $G_{\Bbb R}$ of the complex algebraic group
$G$. Introduce variables $\chi^{\pm}_i=-u^{\pm}_i$. The domain
$u^{\pm}_i >0$ we will call positive domain. The following is
clear.

\begin{prop} Functions $Ch_V(u^+,u^-)$ are positive for
$u^+_i,u^-_i>0$ and
\begin{eqnarray*}
Ch_V(u^+,u^-) &=& Tr_V\left( \prod^r_{j=1}\big(
\frac{u^+_j}{u^-_j}\big)^{h^j} \phi_{i_1}\left(\begin{array}{cc}
1&u^+_{i_1}\\1 & 1+u^+_{i_1}\end{array}\right)\dots
\phi_{i_r}\left(\begin{array}{cc}1 & u^+_{i_r}\\ 1 &
1+u^+_{i_r}\end{array}\right)\right)\\ &=&  Tr_V\left(
\prod^r_{j=1}\big( \frac{u^+_j}{u^-_j}\big)^{h^j}
\phi_{i_1}\left(\begin{array}{cc} 1+u^-_{i_1}&u^-_{i_1}\\1 &
1\end{array}\right)\dots \phi_{i_r}\left(\begin{array}{cc}1+u^-_{i_r} &
u^-_{i_r}\\ 1 & 1\end{array}\right)\right)\\
\end{eqnarray*}
\end{prop}

It is also clear that the map $\a$ is defined globally on positive domain:
\[
\a(u^+_i)=u^-_i \ , \qquad \a(u^-_i)=\frac{(u^-_i)^2}{u^+_i}
\prod_j (1+u^-_j)^{-\B_{ji}} \ .
\]
Let $G_{>0}$ be positive part of $G_{\Bbb R}$ (see
\cite{Lusztig94} and \cite{FominZelevinsky98} for definitions).
For $SL(n)$ the positive part consists of all real unimodular $n\times
n$ matrices with positive principal minors.

\begin{lem} On $G_{>0}$  there exists unique
factorization
\[
g=g_+(g_-)^{-1}
\]
where \ $g_{\pm}^{\pm 1}\in B^{\pm}_{>0}=B^{\pm}\cap G_{>0}$ \ and
\ $\theta(g_+)=\theta^-(g_-)^{-1}$.
\end{lem}

Let $S^{+}_{w,w}$ be the positive Coxeter symplectic leaf of
$G_{\Bbb R}$. It is the connected component of $\varphi^{-1}$ of
the corresponding orbit in $D(G_{\Bbb R})/j({G_{\Bbb R}}_-)$ which
lies in $G_{>0}$. The positive domain described above is
essentially a positive symplectic leaf and thus, on the positive
domain the factorization map $\a$ is the restriction of the
factorization map $g=g_+g_-^{-1}\mapsto {\bar g}=g^{-1}_-g_+$.

\section{The interpolating flow and continuous time nonlinear Toda
  lattices}

\subsection{Interpolating flow}

>From now on we consider the factorization dynamics in positive
real domain. As it was already pointed out the factorization
dynamics on the positive real domain is a graph of a Poisson map. The
trajectory of 
this map is defined recursively as $x(n+1)=x_-(n)^{-1}x_+(n)$ for
$x(n)=x_+(n)x_-(n)^{-1}$.

\begin{prop} The trajectory of the factorization map restricted to the
  positive real domain which starts
at $x(0)$ has the form: $$ x(n) = g_+(n)^{-1}x(0)g_+(n),$$
$g(n)=x(0)^n = g_+(n)g_-(n)^{-1}$.
\end{prop}
\bew\ $x(0)^n=x_+(0)x(1)^{n-1}x_-(0)^{-1}=x_+(0)\dots
x_+(n)x_-(n)^{-1} \dots x_-(0)^{-1}$ shows that
$g_+(n)=x_+(0)\dots x_+(n)$ which quickly leads to the
statement.\eb\bs

This proposition is a discrete analogue of the following theorem of 
Semenov-Tian-Shansky \cite{STS85} for continuous time
systems which describes the trajectories of Hamiltonian systems on
Poisson Lie groups generated by $Ad$-invariant functions. Define
the $\lie$-valued gradient $\nabla f$ of a function $f: G \to
\Real$ by $(\nabla f(g),\eta) := <df(g),(X_\eta)>$ where we write
$X_\eta$ for the left invariant vector field on $G$ corresponding
to $\eta$ and $<\omega,X>$ is the value of the form $\omega$ on
the vector field $X$.

\begin{thm}
Let $H$ be an $Ad_G$-invariant function on $G$.
The trajectory $x(t)$ of the Hamiltonian equations of motion
generated by $H$ is given by $$ x(t) = g_+(t)^{-1} x(0) g_+(t) $$
where $g(t)=\exp (t\nabla H(x(0)) =
g_+(t)g_-(t)^{-1}$.
\end{thm}

Now we are in a position to derive a Hamiltonian flow which
interpolates the factorization dynamics. Obviously, a Hamiltonian
$H_d$ which has a flow whose time $1$ map is given by
factorization as above has to solve the equation $$ g = \exp
(\nabla H_d).$$ Thus, for $g=e^\xi$ and $\xi\in\lie$ we should
have $\xi = \nabla H_d(e^\xi)$ .

\begin{prop}
In a neighborhood of the identity all $Ad_G$-invariant solutions
of the equation
\begin{equation}\label{interpol}
\xi = \nabla H(e^\xi)
\end{equation}
have the form
 $$  H_d(e^\xi) =
\frac{1}{2}(\xi,\xi)+ const. $$
\end{prop}
\bew

 Let  $H$ an $Ad_G$-invariant
solution of the above equation and $\tilde H = H\circ \exp$. Then
$\tilde H$ is $ad_{\mathfrak g}$-invariant and hence $d\tilde
H_\xi(ad_\eta(\xi))=0$ for all $\xi,\eta\in\lie$. By
(\ref{interpol}), $(\xi,\eta) =< dH(e^\xi),X_\eta> = <d\tilde
H(\xi),\eta>$. Here we trivialized the tangent bundle on $G$ by left
translations. Thus, for $\tilde H$ we have the equation
$(\xi,\eta) = d\tilde H|_\xi(\eta)$. Integration yields now the
statement of the proposition.\eb\bs

If $G=SL(n,\Real)$ then $H_d(g)=\frac{1}{2}\mbox{\rm
tr}(\log^2(g))$ in a sufficiently small neighborhood of the
identity \cite{Suris91a}.

The Hamiltonian $H_d$ is quite remarkable since it gives the
so-called classical quantum $R$-matrix
\cite{WeinsteinXu92,Reshet91}. The function $H_d$ is the most
singular part of the quantum $R$-matrix in the appropriate
semi-classical limit \cite{Sklyanin82,Reshet95,Reshet96}. The
map $\a$ generated by time $1$ flow of $H_d$ is the classical
quantum R-matrix in the sense of \cite{WeinsteinXu92} restricted
to the product of Coxeter symplectic leaves and reduced by
hamiltonian reduction.

\subsection{Linearization in a neighborhood of $1$}

Consider the family of diffeomorphisms of ${\Bbb R}^{2n}$ to ${\Bbb R}_+^{2n}$:
\begin{eqnarray*}
\b:(0,1]\x {\Bbb R}^{2n}& \to & {\Bbb R}_+^{2n}\\
\b_{\e}(\pi_i) &=& \e^2 \ e^{\phi_i+\e\pi_i}\\
\b_{\e}(\phi_i)&=& \e^2 \ e^{\phi_i}
\end{eqnarray*}
Here $(\pi_i,\phi_i)$ are coordinates in ${\Bbb R}^{2n}$ such that
$(\b_{\e}(\pi_i),\b_{\e}(\pi_i))$ are the coordinates $(u^+_i,u^-_i)$ which were
used above. Then, assuming a symplectic structure on ${\Bbb R}^{2n}$ s.t.
\begin{equation}\label{2star}
\{\phi_i,\phi_j\}=\{\pi_i,\pi_j\}=0 \ , \qquad
\{\phi_i,\pi_j\}=d_i \B_{ij}
\end{equation}
the maps $\b_{\e}$ are symplectomorphisms.

Define maps $\a_\e:{\Bbb R}^{2n}\to{\Bbb R}^{2n}$, $\e\in (0,1]$ as
$\a_{\e}=\b^{-1}_\e\circ\a\circ\b_{\e}$. They act on coordinates
$(\phi_i,\pi_i)$ as
\begin{eqnarray} \label{ae}
\a_{\e}(\pi_i) &=& \pi_i+\sum^r_{j=1} \B_{ji} \ \frac{1}{\e} \ \ln
(1+\e^2 e^{\phi_j}) \nonumber\\ \a_{\e}(\phi_i) &=&
\phi_i+\e\pi_i+\sum^r_{j=1} \B_{ji} \ \frac{1}{\e} \ \ln (1+\e^2
e^{\phi_j}).
\end{eqnarray}
By construction these maps are symplectomorphisms for (\ref{2star}).

In the limit $\e\to 0$ equation (\ref{ae}) defines a vector
field on $\Bbb R^{2n}$ with coordinates
\begin{eqnarray*}
{\dot\phi_i} &=& \lim_{\e\to 0} \frac{\a_{\e}(\phi_i)-\phi_i}{\e}=
\pi_i \nonumber\\
{\dot\pi_i} &=& \lim_{\e\to 0} \frac{\a_{\e}(\pi_i)-\pi_i}{\e}=
\sum_j \B_{ji} \ e^{\phi_j}
\end{eqnarray*}
This is the Hamiltonian vector field (assuming symplectic structure
(\ref{2star})) generated by the (usual) Toda Hamiltonian
\[
H_{\mbox{Toda}}=\textstyle{\frac 12}(\xi_0,\xi_0)
\]
where
\[
\xi_0=\sum^r_{i=1} (\pi_ih^i+e^{\phi_i} x^-_i+x^-_i) \ .
\]
Thus the family of maps (\ref{ae}) ``retracts" to the Toda
Hamiltonian flow in the neighborhood of the identity.

Equivalently, we have:
\[
\lim_{ n\to\infty}
\a^n_{\e}(\phi,\pi)=(\phi(t),\pi(t))
\]
where $t=n\epsilon$ is fixed and $\phi(t),\pi(t)$ is the Hamiltonian
flow generated by $H_{\mbox{Toda}}$ 
passing through $(\phi,\pi)$ at $t=0$.

It is easy to find the leading terms of the asymptotic expansion of
the integrals 
in the limit $\e\to 0$. Indeed, composing map $\b_{\e}$ with functions
$Ch_V$ and $H_d$ we 
have:
\begin{eqnarray*}
H_V(\phi,\pi)&=&(Ch_V\circ\b_{\e})(\phi,\pi) =
Tr_V(\exp(\xi_{\e}))\\ H_d(\phi,\pi)& =& \textstyle{\frac
12}(\xi_{\e},\xi_{\e})
\end{eqnarray*}
where
\[
\exp(\xi_{\e})=\prod^r_{j=1} \exp(\e\pi_j
h_j)\prod^{\rightarrow}_i \exp(\e e^{\phi_i}x^-_i)\exp(\e x^-_i)
\]
As $\e\to 0$,
\[
\xi_{\e}=\e\xi_0+O(\e^2) \ ,
\]
Thus, for $H_V$ and $H_d$ we have
\begin{eqnarray*}
H_V&=& \dim V\big( 1+\e^2\frac{c_V}{\dim({\frak g})}
H_{\mbox{Toda}} +O(\e^3)\big)\\ H_d  &=& \e^2 H_{\mbox{Toda}}
+O(\e^3)
\end{eqnarray*}
Here we assumed that $V$ is irreducible and $c_V$ is the value of
the Casimir operator action on $V$. Higher Toda Hamiltonians can
be obtained from higher order terms of $\e$-expansion of $H_V$.

\section{Conclusion}

As it was mentioned in the introduction, the main goal of this
paper was systematic derivation of Coxeter-Toda systems from the
symplectic geometry of Poisson Lie groups. Naturally, such
analysis can be done for loop groups as well. The corresponding
models will be affine versions of Coxeter-Toda systems. For $A_n$
root system this will give the relativistic Toda chain first
described by Ruijsenaars \cite{Ruijsenaars90}. In a similar way
one can construct discrete versions of Toda field theories. For
$A_n$- case it has been done in \cite{KashaevReshetikhin97}.

Notice also that somewhat unexpectedly the same Hirota equations
appear as a system of equations for transfer-matrices of some
solvable models in statistical mechanics
\cite{BazhResh90}\cite{KunibaNakanishiSuzuki94}. Although it is clear that the
explanation of this coincidence lies in the theory of $q-W$-algebra
\cite{FrenResh97}, the complete picture is still missing.

The factorization dynamics restricted to other symplectic leaves
will give "nonlinear" Toda-Kostant systems which are related
to general coadjoint orbits.

\section{Acknowledgment} The authors thank
Yuri Suris for helpful discussions. N.R. thanks S. Fomin and A. Zelevinsky for
valuable discussions and the Technische Universit\"at Berlin for the hospitality. 
The research of N.R. was partially supported by the NSF grant
DMS-9603239. T. H., J. K. and N. K. were supported by the  grant
``Discrete integrable systems'' of the Deutsche Akademische
Austauschdienst and by the Sonderforschungsbereich 288
supported by the Deutsche Forschungsgemeinschaft. 

\bibliography{DToda}

\begin{thebibliography}{QNCvdL84}

\bibitem[Adl79]{Adler79}
M.~Adler.
\newblock On a trace functional for formal pseudo-differential operators and
  the symplectic structure of the kdv-type equations.
\newblock {\em Inv. Math.}, 50:219--248, 1979.

\bibitem[Arn89]{Arnold89}
V.~I. Arnold.
\newblock {\em Mathematical Methods of Classical Mechanics, Second Edition}.
\newblock Springer, 1989.

\bibitem[BR90]{BazhResh90}
V.~Bazhanov and N.~Reshetikhin.
\newblock Restricted solid-on-solid models connected with simply laced algebras
  and conformal field theory.
\newblock {\em J.Phys.A}, 23:1477--1492, 1990.

\bibitem[DCKP95]{DeConciniKacProcesi95}
C.~De~Concini, V.G. Kac, and C.~Procesi.
\newblock Some quantum analogues of solvable {Lie} groups.
\newblock In {\em Geometry and analysis. Papers presented at the Bombay
  colloquium, India, January 6--14, 1992}, pages 41--65. Oxford University
  Press, 1995.

\bibitem[DJM82]{DateJimboMiwa82}
F.~Date, M.~Jimbo, and T.~Miwa.
\newblock Method for generating discrete soliton equations {I-IV}.
\newblock {\em J. Phys. Soc. Japan}, 51:4116--4131, 1082.

\bibitem[DLNT86]{DLNT86}
P.~Deift, L.~C. Li, T.~Nanda, and C.~Tomei.
\newblock The {Toda} flow on a generic orbit is integrable.
\newblock {\em Comm. Pure Appl. Math.}, 39:183--232, 1986.

\bibitem[DLT89]{DeiftLiTomei89}
P.~Deift, L.~C. Li, and C.~Tomei.
\newblock Matrix factorization and integrable systems.
\newblock {\em Comm. Pure Appl. Math.}, 42:443--521, 1989.

\bibitem[Dri87]{Drinfeld87}
V.~G. Drinfeld.
\newblock Quantum groups.
\newblock In {\em Proc. Intern. Congress of Math. (Berkeley 1986)}, pages
  798--820. AMS, 1987.

\bibitem[ER97]{FrenResh97}
Frenkel E. and N.~Reshetikhin.
\newblock Deformations of {W}-algebras associated to simple {Lie} algebras.
\newblock {\em from-math-QA-archive}, q-alg/9707012:--, 1997.

\bibitem[FZ98]{FominZelevinsky98}
S.~Fomin and A.~Zelevinsky.
\newblock Totally nonconnegative and oscillatory elements in semisimple groups.
\newblock {\em Preprint}, 1998.

\bibitem[FZ99]{FominZelevinsky99}
S.~Fomin and A.~Zelevinsky.
\newblock Double bruhat cells and total positivity.
\newblock {\em Journal of the AMS}, 12:335--380, 1999.

\bibitem[Hir77]{Hirota77}
R.~Hirota.
\newblock Nonlinear partial difference equations {II}. {Discrete}-time {Toda}
  equation.
\newblock {\em J. Phys. Soc. Japan}, 43(6):2074--2078, 1977.

\bibitem[HL93]{HodgesLevasseur93}
T.~Hodges and T.~Levasseur.
\newblock Primitive ideals of {$C_q[SL(3)]$}.
\newblock {\em Commun. Math. Phys.}, 156:581--605, 1993.

\bibitem[HZ94]{HoferZehnder94}
H.~Hofer and E.~Zehnder.
\newblock {\em Symplectic invariants and Hamiltonian dynamics}.
\newblock Birkhauser Verlag, 1994.

\bibitem[KNS94]{KunibaNakanishiSuzuki94}
A.~Kuniba, T.~Nakanishi, and J.~Suzuki.
\newblock Functional relations in solvable lattice models.
\newblock {\em Intern. Journ. of Modern Physics}, 9:5215--5311, 1994.

\bibitem[Kos79]{Kostant79}
B.~Kostant.
\newblock The solution to a generalized {Toda} lattice and representation
  theory.
\newblock {\em Adv. Math.}, 34:195--338, 1979.

\bibitem[KR97]{KashaevReshetikhin97}
R.~Kashaev and N.~Reshetikhin.
\newblock Affine {Toda} systems as an integrable 3-dimensional quantum field
  theory.
\newblock {\em Comm. Math. Phys.}, 188:251--266, 1997.

\bibitem[KS98]{KorogodskiSoibelman98}
L.~Korogodski and Y.~Soibelman.
\newblock {\em Algebras of Functions on Quantum Groups, Part {I}}.
\newblock American Mathmatical Society, 1998.

\bibitem[Lus94]{Lusztig94}
G.~Lusztig.
\newblock Total positivity in reductive groups.
\newblock In {\em Lie theory and geometry: in honor of Bertram Kostant}.
  Birkhauser, 1994.

\bibitem[LW90]{LuWeinstein90}
J.-H. Lu and A.~Weinstein.
\newblock {Poisson Lie} groups, dressing transformations and {Bruhat}
  decompositions.
\newblock {\em J. Differential Geometry}, 31:501--526, 1990.

\bibitem[MV91]{MoserVeselov91}
J.~Moser and A.~Veselov.
\newblock Discete versions of some classical integtable systems and
  factorization of matrix polynomials.
\newblock {\em Commun. Math. Phys.}, 139:217--243, 1991.

\bibitem[QNCvdL84]{QuispelNijhoffCapelVanderLinden84}
G.~Quispel, F.~Nijhoff, H.~Capel, and J.~van~der Linden.
\newblock Linear integral equations and non-linear differential-difference
  equations.
\newblock {\em Physics A}, 125:344--380, 1984.

\bibitem[Res92]{Reshet91}
N.~Reshetikhin.
\newblock Quasitriangularity of quantum groups and quasi-triangular
  {Hopf}-{Poisson} algebras.
\newblock In {\em AMS Summer Reaserch Institute on Algebras, Groups and Their
  Generalization}, pages 111--133. AMS, 1992.

\bibitem[Res95]{Reshet95}
N.~Reshetikhin.
\newblock Quasitriangularity of quantum groups at roots of 1.
\newblock {\em Commun. Math. Phys.}, 170:79--100, 1995.

\bibitem[Res96]{Reshet96}
N.~Reshetikhin.
\newblock Integrable discrete systems.
\newblock In {\em Quantum Groups and their Appliations in Physics}, pages
  445--487. IOS Press, 1996.

\bibitem[Rui90]{Ruijsenaars90}
S.~Ruijsenaars.
\newblock Relativistic {Toda} systems.
\newblock {\em Commun. Math. Phys.}, 122:217--247, 1990.

\bibitem[Skl82]{Sklyanin82}
E.~Sklyanin.
\newblock On some algebraic structures related to the {Yang}-{Baxter} equation.
\newblock {\em Funct. Anal. and its Appl.}, 16:27--34, 1982.

\bibitem[Soi90]{Soibelman90}
Y.~Soibelman.
\newblock Algebra of functions on a compact quantum group and its
  representations.
\newblock {\em Algebra i Analiz}, 2:193--225, 1990.

\bibitem[STS85]{STS85}
M.~Semenov-Tian-Shansky.
\newblock Dressing transformations and {Poisson} group actions.
\newblock {\em Pub. Res. Inst. Math. Sci. Kyoto Univ.}, 21:1237--1260, 1985.

\bibitem[Sur90]{Suris90}
Y.~Suris.
\newblock Discrete time generalized {Toda} lattices: Complete integrability and
  relation with relativistic {Toda} lattices.
\newblock {\em Phys. Lett. A}, 145:113--119, 1990.

\bibitem[Sur91a]{Suris91a}
Y.~Suris.
\newblock Algebraic structure of discrete-time and relativistic {Toda} x
  lattices.
\newblock {\em Phys. Lett. A}, 156:467--474, 1991.

\bibitem[Sur91b]{Suris91b}
Y.~Suris.
\newblock Generalized {Toda} chains in discrete time.
\newblock {\em Leningrad Math. J.}, 2:339--352, 1991.

\bibitem[Sym82]{Symes82}
W.~Symes.
\newblock The {QR}-algorithm and scattering for the finite non-periodic {Toda}
  lattice.
\newblock {\em Physica}, 4D:275--290, 1982.

\bibitem[Tod88]{Toda88}
M.~Toda.
\newblock {\em Theory of nonlinear lattices}.
\newblock Springer, 1988.

\bibitem[Ves91]{Veselov91}
A.~P. Veselov.
\newblock Integrable maps.
\newblock {\em Russ. Math. Surv.}, 46:3--45, 1991.

\bibitem[WX92]{WeinsteinXu92}
A.~Weinstein and P.~Xu.
\newblock Classical solutions to the quantum {Yang-Baxter} equation".
\newblock {\em Commun. Math. Phys.}, 143:309--344, 1992.

\end{thebibliography}
\end{document}